\documentclass{tMPH2e}
\usepackage{graphicx}
\usepackage{mathptmx}
\usepackage{amsmath}
\usepackage{amssymb}
\renewcommand{\H}{\text{H}}
\renewcommand{\S}{\text{S}}
\newcommand{\D}{\text{D}}
\newcommand{\F}{\text{F}}
\newcommand{\E}{\text{E}}
\allowdisplaybreaks

\begin{document}
\doi{0}
\issn{0}
\issnp{0}
\jvol{0}
\jnum{0}
\jyear{2015}
\jmonth{10 January}

\title{Lattice-Fluid Models derived from Density Functional Theory}
\author{Stephan Korden$^\dagger$
\thanks{$^\dagger$ Email: stephan.korden@rwth-aachen.de}\\\vspace{6pt}
{\em{Institute of Technical Thermodynamics, RWTH Aachen University,\\ 
Schinkelstra\ss e 8, 52062 Aachen, Germany}}\\\vspace{6pt}
\received{Version: \today}}
\maketitle

\begin{abstract}
In the current article, we rederive the lattice-fluid excess models UNIQUAC, UNIFAC, and COSMO-RS from a continuum 
functional. The calculation explains the missing dependence on the particle geometry and how to include the Coulomb 
interaction, problems that are common to all three models. It is then shown that the Wilson ansatz, used in UNIQUAC and 
UNIFAC to minimize the grand potential, is not a physically valid solution of the Euler-Lagrange equation. A consistent 
approach is the Larsen-Rasmussen equation, which forms the foundation of COSMO-RS. We then analyze the various 
approximation methods and interpret them in the framework of a molecular density functional.
\end{abstract}
\section{Introduction}\label{sec:intro}
A fundamental problem of density functional theory (DFT) is the construction of the grand-canonical potential, which in 
soft-matter physics has to cover the wide spectrum of interparticle potentials ranging from the strongly repulsive 
inner domains of molecules to the Coulomb interaction of ions \cite{mcdonald,evans-dft}. A common strategy is therefore 
the separation of scales and to approximate the repulsive inner core by a hard-particle potential. Assuming that the 
residual interactions are weak, this reduces the construction of the grand potential to the perturbative expansion of 
the free energy and the derivation of the distribution functionals for hard particles. The latter have been derived 
recently, which leaves us to identify a suitable representation for the molecular functional \cite{korden-2, korden-4, 
korden-6}.

For practical applications, it is necessary that the minimization of the grand potential remains calculational 
manageable. We therefore require the functional to be of low perturbation order, valid for short and long-range 
interactions, to allow further approximations in the hard-particle functionals, and to be numerically simple to 
integrate. It is not easy to meet all these criteria. But a good starting point is the analysis of lattice-fluid excess 
theories, especially the well established universal quasi-chemical model (UNIQUAC), its extension to functional  
activity coefficients (UNIFAC), and the conductor-like screening model for real solvents (COSMO-RS), which are simple 
but fully operational density functionals that derive the liquid-liquid equilibrium by minimizing the grand potential 
\cite{prigog1,praus4,klamt-0}.

Lattice theories were among the earliest models in solid-state physics, best known for the Ising model and its solution 
in one and two dimensions \cite{gub-rev,hill,baxter2} . But they were also applied to the liquid state by Flory, 
Huggins, Staverman, Guggenheim, et al., until experimental results proved the more random structure of the particle 
ordering \cite{flory1,flory2,huggins,gug1}. As a result, they lost its popularity in physics, but remained in use in 
biology, chemistry, and engineering, where they have been extended by Prausnitz, Abrams, and Maurer to the UNIQUAC, 
UNIFAC models and by Klamt to COSMO-RS, with the alternative implementation of COSMO-SAC by Lin and Sandler 
\cite{praus1, praus2, praus3, klamt-0, klamt-1, klamt-2, sac_1, sac_2}. Together with the group-contribution 
approximation, they successfully predict the liquid-liquid equilibria for a large class of molecules using a simplified 
interaction potential for their chemical groups or surface charges.

A major disadvantage is their limitation to a fixed value of density and pressure as a consequence of the lattice
representation of the molecules by linear chains of lattice cells. This allows to distinguish individual particles by 
their positions, which results in a wrong combinatorial factor of the free energy \cite{gub-rev}. The lattice 
functional itself is therefore ill defined. But the incorrect part cancels in the excess energy of a mixture with
particles of maximal packing fraction. This partly corrects the error of the lattice ansatz, but also shows that any 
improvement of the model is only possible in the framework of DFT.

Sec.~\ref{sec:pert} begins with a general discussion of the free-energy representations and their respective 
perturbative expansions. It is then shown in Sec.~\ref{sec:mdft} that the lattice models originate from the dual 
functional. We analyze their solutions of the Euler-Lagrange equation and reinterpret the underlying approximations in 
terms of a continuum molecular functional.
\section{The dual Free-Energy Functional}\label{sec:pert}
The Hohenberg, Kohn, Mermin theorem proves the unique mutual relationship between interaction and grand-canonical 
potential \cite{mcdonald}, forming the foundation of DFT, as it guarantees that different representations of the 
functional determine the same thermodynamic ground state. It thus constraints the number of alternative representations 
for a given interaction potential, which have to be related by similarity transformations. Ignoring mappings that 
correspond to internal symmetries of the potential or result in contributions which cancel in the Euler-Lagrange 
equations, the only nontrivial symmetry is the Legendre transformation of the grand potential $\Omega$, exchanging its 
canonically conjugate variables. For the simplest case of pair interactions, it replaces the potential $\phi_{ij}$ by 
its dual pair-correlation functional $g_{ij}$, defined by
\begin{equation}\label{var-2}
\frac{\delta \Omega}{\delta \phi_{ij}} = \frac{1}{2} \rho_i \rho_j g_{ij}\;.
\end{equation}
This shows that $\Omega(\phi_{ij})$ has only two representations, either as the free-energy functional 
$\Omega^\F(\phi_{ij})$ or its Legendre-dual counterpart $\Omega^\D(g_{ij})$.

Most molecular functionals use $\Omega^\F$ as the starting point, as its perturbation expansion in r-particle densities 
$\rho_{i_1\ldots i_r}$ is algebraically well understood \cite{mcdonald}. But it will be shown in the next section that 
the lattice models derive from the dual functional $\Omega^\D$, whose analytic form is more complex and the 
perturbation expansion of $g_2$ does not result in either the direct or the distribution functionals. For comparison, 
we will derive both representations, analyze their respective perturbation expansions, and discuss their advantages and 
limitations.

Beginning with the free-energy representation, the functional $\Omega^\F( \beta, \mu_i, \phi_i^\text{ex}| \rho_i, 
\phi_{ij})$ of the particle density $\rho_i$ depends on the inverse temperature $\beta = 1/k_\text{B}T$, chemical 
potential $\mu_i$, and external potential $\phi_i^\text{ex}$ for a mixture of $i=1,\ldots, M$ compounds. It is an 
integral
\begin{equation}\label{basic}
\beta\Omega^\F = \sum_{i=1}^M\int \big[ \rho_i(\ln{(\rho_i\Lambda_i)}-1) - \beta\rho_i(\mu_i - \phi_i^\text{ex}) \big] 
\,d\gamma_i - c_0(\rho_i)
\end{equation}
over the positions and orientations $\gamma_i \in \mathbb{R}^n \times \text{SO}(n)$ of the $n$-dimensional Euclidean 
space and depending on the thermal wavelength $\Lambda_i$ and direct correlation functional $c_0(\rho_i)$.

Its perturbation expansion in the potential $\phi = \phi^\H + \phi^\S$ of hard-particle $\phi^\H$ and soft interaction 
$\phi^\S$ is a formal Taylor series of the logarithm of the grand canonical partition integral, whose first and second 
order in the Mayer functions $f^\S_2$ have the form
\begin{equation}\label{gc-pert}
\begin{split}
\beta \Omega^\F = \beta \Omega_\H^\F & - \frac{1}{2}\int \rho_{i_1i_2}^\H f_{i_1i_2}^\S\,d\gamma_{i_1i_2} - 
\frac{1}{2}\int \rho_{i_1i_2i_3}^\H f_{i_1i_2}^\S f_{i_2i_3}^\S\,d\gamma_{i_1i_2i_3}\\
&\qquad \qquad  - \frac{1}{8}\int(\rho_{i_1i_2i_3i_4}^\H - \rho_{i_1i_2}^\H \rho_{i_3i_4}^\H)f_{i_1i_2}^\S 
f_{i_3i_4}^\S \, d\gamma_{i_1i_2i_3i_4} - \ldots\;,
\end{split}
\end{equation}
where a sum over paired indices is implied \cite{mcdonald}. Higher order terms are readily obtained using diagrammatic 
techniques \cite{korden-6}, where each product $[f_2^\S]^m$ couples to a homogeneous polynomial of r-particle densities 
of order $2\leq r \leq 2m$, integrated over the $(m-1)n(n+1)/2$ coordinates of position and orientation. The rapid 
increase in the dimensionality of the integrals effectively limits the perturbative approach to the first or second 
order. 

The expansion (\ref{gc-pert}) requires the particles to interact by pair potentials. But the same approach also applies 
to irreducible m-particle interactions, when the fully $f^\S_2$-bonded subdiagrams $[f_2^\S]^m$ are replaced by the 
Mayer function $f_m^\S$. For the 3-particle interaction $\phi_{i_1i_2i_3}$, e.g., this adds the leading correction
\begin{equation}\label{3-part}
\beta\Omega^\F = \beta\Omega^\F(\phi_{ij}) + \frac{1}{6} \int \rho_{i_1i_2i_3}^\H f_{i_1i_2i_3}^\S\,d\gamma_{i_1i_2i_3} 
+ \ldots 
\end{equation}
to the 2-particle functional $\Omega^\F(\phi_{ij})$.

The second representation is the dual functional $\Omega^\D$. First derived by Morita and Hiroike using diagrammatic 
techniques \cite{morita1,morita2,morita3,morita4,baxter3}, it replaces $\phi_{ij}$ by its canonically conjugate 
variable $g_{ij}$. To perform the Legendre transformation, we integrate (\ref{var-2}) over $\delta\phi_{ij}$
\begin{equation}\label{part-trafo}
\Omega = \Omega_{\text{kin}} + \frac{1}{2}\int \rho_i\rho_j g_{ij}\phi_{ij}\,d\gamma_{ij} - \frac{1}{2}\int 
\rho_i\rho_j \phi_{ij} \delta g_{ij}\, d\gamma_{ij}\;.
\end{equation}
To complete the integration over $\delta g_{ij}$, Morita and Hiroike derive the self-consistent closure relation 
between $\phi_{ij}$ and $g_{ij}$ \cite{morita3, mcdonald}:
\begin{equation}\label{scf}
\ln{(g_{ij})} = -\beta \phi_{ij} + d_{ij} + h_{ij} - c_{ij}\;,
\end{equation}
introducing the bridge functional $d_{ij}$ of 2-path connected clusters, the pair correlation $h_{ij}=g_{ij}-1$, and 
the 2-particle direct correlation functionals $c_{ij}$. To eliminate the remaining dependence on the free-energy 
representation, $c_{ij}$ is then substituted using the Ornstein-Zernike equation
\begin{equation}
\begin{split}
c_{ij} - h_{ij} & = - \int \rho_k\,h_{ik}\,c_{kj}\, d\gamma_k
= \sum_{n=1}^\infty \int (-1)^n \rho_{k_1}\ldots \rho_{k_n}\, h_{ik_1} \ldots h_{k_nj}\,d\gamma_{k_1\ldots k_n}\;.
\end{split}
\end{equation}
Inserted into (\ref{part-trafo}), they form the infinite sum over h-bonded ring integrals
\begin{equation}\label{ring}
\int \rho_i\rho_j ( c_{ij} - h_{ij})\,\delta h_{ij} \,d\gamma_{ij} = \sum_{n=3}^\infty \int \frac{(-1)^n}{n} 
\rho_{k_1}\ldots \rho_{k_n}\, h_{k_1k_2}\ldots h_{k_nk_1}\,d\gamma_{k_1\ldots k_n}
\end{equation}
in the final representation of the dual grand-canonical potential \cite{morita3,baxter3}:
\begin{align}
\beta\Omega^\D\; & = \;\sum_i \int \rho_i \ln{(\rho_i\Lambda_i)} - \rho_i -\beta \mu_i\rho_i \, d\gamma_i
+ \frac{1}{2}\sum_{ij} \int \rho_i\rho_j\big( g_{ij}\ln{(g_{ij})} - g_{ij} + 1 \big)\, d\gamma_{ij}\nonumber\\
&\quad + \frac{\beta}{2}\sum_{ij} \int \rho_i\rho_j\, g_{ij}\, \phi_{ij}\, d\gamma_{ij}
+ \frac{1}{2}\sum_{n=3}^\infty \int \frac{(-1)^n}{n} \rho_{k_1}\ldots \rho_{k_n}\, h_{k_1k_2} \ldots 
h_{k_nk_1}\,d\gamma_{k_1\ldots k_n}\label{mh}\\
&\quad -\frac{1}{2} \int \rho_i\rho_j d_{ij}\,\delta g_{ij}\, d\gamma_{ij}\;,\nonumber
\end{align}
where the integration constant $+1$ in the second integral has been chosen to reproduce the ideal gas in the limit 
$\phi_{ij} \to 0$. 

Compared to the free energy representation (\ref{basic}), the analytic structure of the dual functional is considerably
more complex, while containing exactly the same information for pairwise interacting particles. A common simplification 
is therefore to set $d_{ij}=0$ and to use either the Percus-Yevick (PY) or the hypernetted chain approximation (HNC) 
for the closure relation (\ref{scf}) 
\begin{equation}\label{py-hnc}
\begin{split}
\text{PY}\;\;:& \quad g_{ij}\exp{(\beta \phi_{ij})}=\exp{(d_{ij}+h_{ij}-c_{ij})}\approx 1 + h_{ij} - c_{ij}\\ 
\text{HNC}:& \qquad\;\; \ln{(g_{ij})} \quad \approx -\beta \phi_{ij} + h_{ij} - c_{ij}\;.
\end{split}
\end{equation}
In combination with the Ornstein-Zernike equation \cite{mcdonald}, they provide easier to solve self-consistent 
integral equations for $h_2$ and $c_2$. 

Probably the best known example is the PY approximation for hard spheres and its solution for $g_2$ developed by 
Wertheim, Thiele, and Baxter \cite{wertheim-py1,wertheim-py2,thiele,baxter}. Another example is the Coulomb potential 
$\phi = q^2/r$ for point-particles of charge $\pm q$. Its slow radial decline allows the long-range approximation $c_2 
\approx -\beta \phi$, for which the HNC equation can be solved, using the Fourier transformation $\hat c_2=F(c_2)$ to 
decouple the Ornstein-Zernike equation
\begin{equation}\label{debye}
\ln{(g_2)} = -\beta\phi + h_2 - c_2 \approx h_2 = F^{-1}\Big(\frac{\hat c_2}{1 - \rho \hat c_2}\Big) 
= -\beta \frac{q^2}{r} \exp{(-k_D r)}\;.
\end{equation}
This result reproduces the characteristic Debye-H{\"u}ckel screening for an ionic liquid of wavenumber $k_D = (4\pi 
\beta \rho q^2)^{1/2}$ and, together with the infinite sum over the ring integrals
\begin{equation}\label{dh}
\int\rho (\hat c_2 -\hat h_2) \delta\hat h_2\,d\hat\gamma = \rho\hat h_2 -\frac{1}{2}\rho^2 \hat h^2_2 - \ln{(1+ 
\rho\hat h_2)}\;,
\end{equation}
yields the Debye-H{\"u}ckel functional for charged spherical particles \cite{mcdonald,pitzer}.

The calculation illustrates how the combination of Ornstein-Zernike and closure equation (\ref{scf}) improves the low 
order approximation. Actually, it is an example of a duality transformation that inverts the length scales by 
exchanging a pair of canonically conjugate variables, mapping the perturbative sector of one functional to the 
non-perturbative of its dual. This shows that the two representations $\Omega^\F$ and $\Omega^\D$, although equivalent 
in their total information, have different application ranges when the perturbation series are restricted to a finite 
order. 

\begin{figure}
\centering
\includegraphics[width=2.9cm,angle=-90]{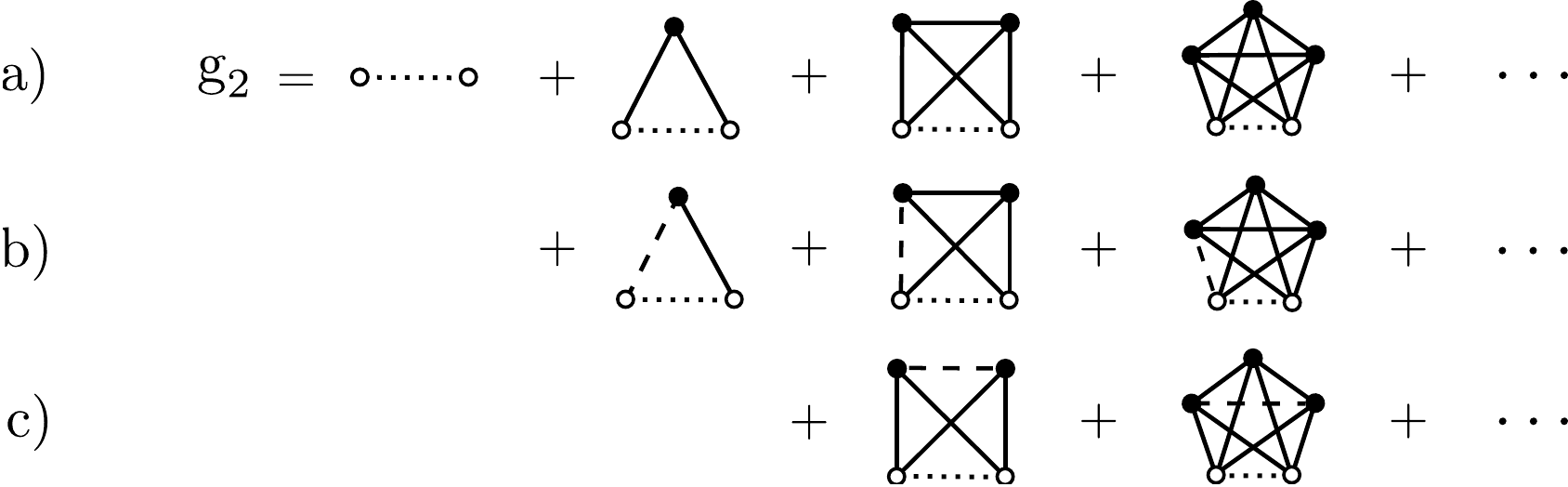}
\caption{The perturbative expansion of the pair-correlation functional $g_2$ written in Mayer diagrams: solid lines 
indicate $f^\H$-bonds, dashed lines correspond to $F$-bonds, and dotted ones represent the product $e_2^\S e_2^\H$ of 
hard- and soft- Boltzmann functions. The main contribution derives from the fully bonded diagrams a) with no internal 
$F$-bond $g_2|_{2,0}$, b) one $F$-bond linked to a rooted node $g_2|_{2,1}$, and c) one internal $F$-bond $g_2|_{4,1}$.}
\label{fig:pert-diag}
\end{figure}

Contrary to the expansion (\ref{gc-pert}), the perturbation theory for $\Omega^\D$ is an expansion of $g_2$ in the 
soft correction term $F_{ij} = e_{ij}^\H f_{ij}^\S$ of the Mayer $f_{ij} = f_{ij}^\H + F_{ij}$ and Boltzmann $e_{ij}$ 
functions. Its lowest order diagrams are shown in Fig.~\ref{fig:pert-diag}, illustrating the successive replacement of 
$f^\H$-bonds by $F$-functions. Its corresponding substitution in the functional is generated by the formal derivative
\begin{equation}\label{pert-g}
\begin{split}
g_{ij} &= e_{ij}^\S g_{ij}^\H + e_{ij}^\S \int \frac{\delta g_{ij}^\H}{\delta f_{ik}^\H}\, F_{ik}\, d\gamma_k
+ \frac{1}{2} e_{ij}^\S \int \frac{\delta g_{ij}^\H}{\delta f_{k_1k_2}^\H}\, F_{k_1k_2}\, d\gamma_{k_1k_2} + \ldots\\
& \approx g_{ij}|_{2,0} + g_{ij}|_{2,1} + g_{ij}|_{4,1} + \ldots\;,
\end{split}
\end{equation}
whose contributions can be further approximated by a series of correlations $g_2|_{n,k}$ of $n$ hard-particle 
intersection centers and $k$ internal $F$-bonds.
Using the notation of \cite{korden-4,korden-6}, the functionals of lowest intersection order and with at most one 
internal $F$-bond have intersection diagrams of the form:
\begin{equation}
\begin{split}
\widetilde\Lambda_{2,1}^{(2)}\;&:\quad e_{i_1i_2} + e_{i_1i_2} \gamma_a^{i_1i_3\ldots i_n} \gamma_b^{i_2i_3\ldots i_n}\\
\widetilde\Lambda_{3,1}^{(2)}\;&:\quad e_{i_1i_2}F_{i_2i_3}\gamma_a^{i_1i_4\ldots i_n}\gamma_b^{i_2i_3i_4\ldots i_n}\\
\widetilde\Lambda_{4,1}^{(2)}\;&:\quad e_{i_1i_2}F_{i_3i_4}\gamma_a^{i_1i_3i_5\ldots i_n}\gamma_b^{i_1i_4i_5\ldots i_n}
\gamma_c^{i_2i_3i_5\ldots i_n}\gamma_d^{i_2i_4i_5\ldots i_n}\;,
\end{split}
\end{equation}
whose resummation yields the functionals
\begin{equation}
\begin{split}
g_{i_1i_2}|_{2,0} &= e_{i_1i_2}^\S e^\H_{i_1i_2}(\delta_{AI} + \mathcal{D}_a\mathcal{D}_b 
\frac{w_a^{i_1}w_b^{i_2}}{1-x_{ab}})\\ 
g_{i_1i_2}|_{2,1} &= - e^\S_{i_1i_2} e^\H_{i_1i_2} \mathcal{D}_a\mathcal{D}_b \int \frac{w_a^{i_1}w_b^{i_2}
F_{i_1i_3} w_b^{i_3}\rho_{i_3}}{(1-x_{ab})^2}\,d\gamma_{i_3}\\
g_{i_1i_2}|_{4,1} &= - e^\S_{i_1i_2} e_{i_1i_2} \mathcal{D}_a\mathcal{D}_b\mathcal{D}_c\mathcal{D}_d 
\frac{(w_a^{i_1}w_b^{i_1})
(w_c^{i_2}w_d^{i_2}) y_{acbd}}{(1-x_{abcd})^3}\;,
\end{split}
\end{equation}
with the correlation integral of the inner bond
\begin{equation}
y_{abcd} = \int F_{i_3i_4} (w_a^{i_3}w_c^{i_3}\rho_{i_3}) (w_b^{i_4}w_d^{i_4}\rho_{i_4})\,d\gamma_{i_3i_4}\;.
\end{equation}

This example shows that the representation in intersection centers can also be extended to the perturbative expansion 
of $g_2$, completing previous results for the direct and distribution functionals \cite{korden-4,korden-6}. But for 
many applications it will be sufficient to restrict the series to the first order in the hard-particle correlations and 
to use $g_2 \approx e_2^\S g_2^\H|_{2,0}$. This approximation has the additional advantage to satisfy $g_2(\vec r) \geq 
0$ and $g_2(r\to\infty)=1$ for any $\vec r \in \mathbb{R}^n$. These two conditions are easily tested and provide an 
important advantage over $\Omega^\F$, where no such constraints exists. Because $-1\leq f_2^\S < \infty$ and the 
discontinuity of $g_2^\H$ at particle contact, the integral of its first approximation order $f_2^\S g_2^\H$ is 
indefinite and very sensitive to changes of the hard-particle geometry and soft-interaction potential. Another 
disadvantage is the dependence of (\ref{gc-pert}) on $c_0^\H$ and $g_2^\H$, which requires the redundant calculation of 
two functionals with the same information content.

In summary, $\Omega^\D$ depends only on one class of correlation functionals, allows a simple consistency test for its 
perturbative corrections, and applies for short- and long-range interactions alike. A disadvantage, however, is its 
limitation to pair interactions. But as higher-order irreducible m-particle potentials are often very short-ranged, 
they can be coupled perturbatively using the expansion (\ref{3-part}). Thus, despite its complex structure, the dual 
functional is the preferred ansatz for the perturbative construction of a molecular model, which will be further 
supported in the next section, where we make contact with the lattice theories. 
\section{Lattice-Fluid Models derived from the Dual Functional}\label{sec:mdft}
Lattice models for fluids use a discretized Euclidean space, with molecules represented by linear chains of cells, 
reducing the configuration integral to a sum over all allowed particle insertions \cite{gug1,prigog1,praus4,baxter2}. To 
simplify the derivation, two additional assumptions are made: 1. molecules are closely stacked, i.e. the packing 
fraction for all systems is $\eta = 1$, and 2. interactions only occur between next neighbors. 

A mixture of $N=\sum_i N_i$ particles with $i=1, \ldots, M$ compounds is therefore independent of volume effects, from 
which follows that the excess free energy of mixing
\begin{equation}\label{mix}
F^\E(\{x_i\}) = F(\{x_i\}) - \sum_i x_i\, F_i(x_i=1)
\end{equation}
is a function of temperature and the molar fractions $x_i=N_i/N$. 

To derive the lattice model from the functional $\Omega^\D$, we have to interpret its two assumptions in terms of the 
continuum formulation. The first constraint of close packing is readily implemented for a mixture of constant densities 
$\rho_i = N_i/V$ and their pure-compound systems $\hat \rho_i = N_i/V_i$ with partial volumes $V_i = x_i V$ and 
molecular volumes $v_i$:
\begin{equation}\label{eta}
1 = \eta = \sum_k \rho_k v_k = \hat \rho_i v_i = \hat \eta_i\;.
\end{equation}
The second constraint reduces the correlation length of the pair-distribution function to its next neighbors. In a 
first step, we therefore neglect all $g_2$ in the functional (\ref{mh}) beyond the leading order
\begin{equation}\label{reducf}
\beta F = \sum_i\int \rho_i\ln{(\rho_i\Lambda_i)} - \rho_i\,d\gamma_i +\frac{1}{2}\sum_{ij}\int \rho_i\rho_j 
\big(g_{ij}\ln{(g_{ij})} - g_{ij} +1 + \beta g_{ij}\phi_{ij}) \, d\gamma_{ij}\;,
\end{equation}
thus removing the bridge and ring integrals responsible for the Debye-H{\"u}ckel screening. Next, we restrict the 
spacial range of $g_2$ to the first particle shell $\Lambda_{ij}$, which comes closest to the idea of next-neighbor 
correlations between cell elements. Introducing the definitions
\begin{equation}\label{defz}
z_{ij} = \rho_j g_{ij}|_{\Lambda_{ij}}\;,\quad 
z_i = \sum_j \int_{\Lambda_{ij}} \rho_j g_{ij} \,d\gamma_j \;,\quad z = \frac{1}{2}\sum_{ij} \int_{\Lambda_{ij}} \rho_i 
\rho_j g_{ij}\,d\gamma_{ij}\;,
\end{equation}
the pair density and correlation functions can be rewritten
\begin{equation}
\begin{split}\label{ident}
\rho_i\rho_j g_{ij}|_{\Lambda_{ij}} = \frac{z_{ij}}{z_j}\,\frac{\rho_j z_j }{z}\, 
z = \theta_{ij}\,\theta_j\, z\;& ,\quad 
g_{ij}|_{\Lambda_{ij}} = \frac{z_{ij}}{z_j}\,\frac{z_j}{\rho_i} = \frac{\theta_{ij}}{\theta_i} 
\frac{z_i\,z_j}{z}\\
z_i = \frac{\rho_i z_i}{z}\frac{z}{\rho_i} = 
\frac{\theta_i}{\phi_i}\frac{z}{\rho}\;& ,\quad 
\phi_i = \frac{\rho_i}{\hat \rho_i} = \frac{\eta_i}{\eta}
\end{split}
\end{equation}
in terms of the local coordination number $\theta_{ij}=z_{ij}/z_j$, surface fraction $\theta_i=\rho_iz_i/z$, and the
volume fraction $\phi_i$ of lattice theories. These variables are not independent, but related by the permutation 
symmetry $\rho_{ij} = \rho_{ji}$ and normalization
\begin{equation}\label{norm}
\theta_{ij}\theta_j = \theta_{ji}\theta_i\;, \qquad \sum_{i} \int_{\Lambda_{ij}} \theta_{ij}\,d\gamma_i = 1\;, \qquad 
\sum_i \int_{\Lambda_{ij}} \theta_i \, d\gamma_i = 1\;.
\end{equation}
Using these identities, the continuum functional (\ref{reducf}) can now be written in the basis of lattice variables.

Beginning with the potential energy
\begin{equation}\label{pot}
\frac{1}{2} \sum_{ij} \int_{V\times V} \rho_i \rho_j g_{ij}\phi_{ij}\,d\gamma_{ij} =  
\frac{1}{2} N \sum_{ij} \int_{\Lambda_{ij}} \rho_i\rho_j g_{ij}\phi_{ij}\,d\gamma_{ij} = \frac{1}{2}\, zN 
\sum_{ij}\theta_{ij}\,\theta_j\, 
\epsilon_{ij}\;,
\end{equation}
the integration over $V\times V$ separates into a sum over $N/2$ particle pairs of volume $\Lambda_{ij}$, while the 
potential $\phi_{ij}$ is approximated by the constant energy parameter $\epsilon_{ij}$ of neighboring cells molecules. 
The same transformation also applies to the logarithmic term of (\ref{reducf})
\begin{equation}\label{kin-g}
\begin{split}
& \frac{1}{2}\sum_{ij}\int_{V\times V} \rho_i\rho_j g_{ij}\ln{(g_{ij})}\,d\gamma_{ij} 
= \frac{1}{2} \,zN \sum_{ij} \theta_{ij} \theta_j\, \ln{\Big( \frac{\theta_{ij}}{\theta_i}\, \frac{z_i\,z_j}{z}\Big)}\\
&\qquad\qquad = \frac{1}{2}\,zN\,\sum_{ij}\theta_{ij}\,\theta_j \ln{\Big( \frac{\theta_{ij}}{\theta_i}\Big)} + 
zN\,\sum_i \theta_i \ln{\Big( \frac{\theta_i}{\phi_i}\Big)} + \frac{1}{2}\,zN\,\ln{\Big(\frac{z}{\rho^2}\Big)}\;,
\end{split}
\end{equation}
whose constant contribution cancels in the excess free energy (\ref{mix}). The same applies to the linear term
\begin{equation}\label{lin-g}
\frac{1}{2}\sum_{ij}\int_{V\times V} \rho_i\rho_j(g_{ij} - 1)\,d\gamma_{ij} = \frac{1}{2}N(2z - N)\;.
\end{equation}

Slightly more complicated is the transformation of the kinetic energy, as the integration over the configuration space 
$\Lambda_{ij}$ effectively reduces the number of independently moving molecules. The excess kinetic energy of unpaired 
particles
\begin{equation}\label{f1}
\begin{split}
\beta F^\text{E,1}_\text{kin} & = \sum_i\int_V (\rho_i\ln{(\rho_i\Lambda_i)} - \rho_i) \,d\gamma_i 
- \sum_i \int_{V_i} (\hat\rho_i\ln{(\hat\rho_i\Lambda_i)} - \hat\rho_i) \, d\gamma_i\\
&= \sum_i N_i \ln{\Big(\frac{\rho_i}{\hat\rho_i}\Big)}
 = N\sum_i x_i \ln{(\phi_i)}\;,
\end{split}
\end{equation}
has to be corrected by the kinetic energy of the particle pairs. To determine their contribution, one has to observe 
that the translational and rotational degrees of freedom of one particle is bound to the second particle of the cluster. 
We therefore subtract for each pair the kinetic energy of one particle. The amount of energy bound by the density of 
$\rho_i z_i/2$ particle pairs is determined by the difference:
\begin{equation}
\begin{split}
& \frac{1}{2} \sum_i \int_V (\rho_iz_i \ln{(\rho_i z_i\Lambda_i)} - \rho_i z_i) \,d\gamma_i
- \frac{1}{2}\sum_i\int_V z_i(\rho_i \ln{(\rho_i\Lambda_i)} - \rho_i)\,d\gamma_i\\
&\hspace{9em}  = \frac{1}{2}\sum_i \int_V \rho_i z_i \ln{(z_i)}\;.
\end{split}
\end{equation}
From this derives the excess kinetic energy stored by the clusters
\begin{equation}
\begin{split}
\beta F^\text{E,2}_\text{kin} & = \frac{1}{2} \sum_i \int_V \rho_i z_i \ln{(z_i)} \,d\gamma_i
- \frac{1}{2}\sum_i\int_{V_i} \hat \rho_i \hat z_i \ln{(\hat z_i)}\,d\gamma_i\\
&= \frac{1}{2}zN\sum_i \theta_i \ln{\Big(\frac{\theta_i}{\phi_i}\frac{z}{\rho}\Big)} 
- \frac{1}{2}zN \ln{\Big(\frac{z}{\rho}\Big)}
= \frac{1}{2}zN\sum_i \theta_i \ln{\Big(\frac{\theta_i}{\phi_i}\Big)}\;.
\end{split}
\end{equation}
Subtracting this result from the excess kinetic energy of the free particles (\ref{f1}), yields the effective kinetic 
energy of the lattice fluid
\begin{equation}
\beta F^\E_\text{kin} = \beta F^\text{E,1}_\text{kin} - \beta F^\text{E,2}_\text{kin} 
= N\sum_i x_i \ln{(\phi_i)} - \frac{1}{2}zN\sum_i \theta_i \ln{\Big(\frac{\theta_i}{\phi_i}\Big)}\;.
\end{equation}

Combining the previous results (\ref{pot}), (\ref{lin-g}), (\ref{kin-g}) with the identities $\hat\theta_{ii} = 1$, 
$\hat\theta_i = 1$ for the pure compounds and canceling constant contributions, we finally arrive at the excess free 
energy of the lattice liquid
\begin{equation}\label{uni}
\beta F^\E/N  = \sum_i x_i\ln{(\phi_i)} + \frac{z}{2}\sum_i\theta_i \ln{\Big(\frac{\theta_i}{\phi_i}\Big)}
+ \frac{z}{2}\sum_{ij} \theta_{ij}\theta_j \Big[ \ln{\Big(\frac{\theta_{ij}}{\theta_i}\Big)}
+ \beta (\epsilon_{ij} - \epsilon_{jj}) \Big]\;,
\end{equation}
whose first two terms correspond to the results from Flory-Huggins and Staverman-Guggenheim \cite{flory1, flory2, 
huggins, gug1}. The corresponding grand-canonical excess functional follows by adding the excess chemical potential of 
paired particles
\begin{equation}\label{chem}
\Omega^\E(\theta_{ij}) = F^\E(\theta_{ij}) - zN \sum_i \theta_i \mu_i\;.
\end{equation}

\begin{figure}
\centering
\includegraphics[width=3.5cm,angle=-90]{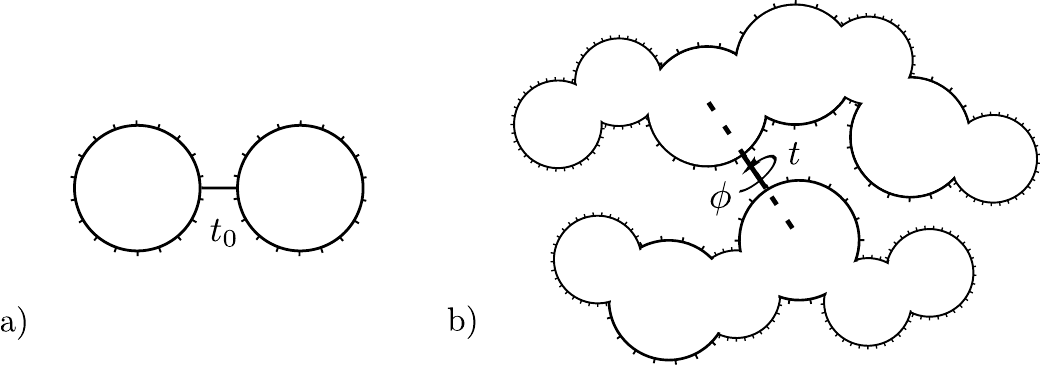}
\caption{Comparing the interaction models of COSMO-RS and the molecular functional: a) The COSMO-RS model maps the 
surfaces of the molecules to unit spheres, with partial charges interacting over a fixed distance $t_0$. b) The 
molecular functional determines the pair-correlation functional, coupling the hard-particle geometry to the soft 
interaction of partial charges. The grand potential is the integral over all segment-segment combinations, distances 
$t$, and axial rotations $\phi$.}
\label{fig:segs}
\end{figure}

The mixtures are now uniquely determined by the four sets of variables $\theta_{ij}$, $\theta_i$, $\phi_i$, $z$, and 
the constraints (\ref{norm}). But only $\theta_{ij}$ is fixed by the Euler-Lagrange equation of $\Omega^\E$. The 
remaining variables still need to be determined from their definitions (\ref{defz}), (\ref{ident}), and the assumptions 
of the lattice model. The molecules, e.g., are flexible, linear chains of cells without self-intersection. Their 
specific shape is therefore undefined, but the volumes $v_i$ and surfaces $a_i$ are constant and the contact 
probability independent of positions and orientations, thus corresponding to $g^\H_{ij}|_{\Lambda_{ij}} \approx c$. 
The hard-particle pair-correlation is then a function 
\begin{equation}\label{glat}
g^\H_{ij}(t)|_{\Lambda_{ij}} = c e_{ij}(t) \delta(t)
\end{equation}
for particles whose surfaces are separated by the distance $t=0$, as shown in Fig.~\ref{fig:segs}a).

To calculate the coordination numbers (\ref{defz}), we use the representation from App.~\ref{sec:app} for the 
integral measure $d\gamma_{ij}$ of two particles with principal curvatures $\kappa_\alpha^{(i)}$ at a distance $t=0$ 
and rotation angle $0\leq \phi < 2\pi$. Expanding the determinant (\ref{g-12})
\begin{equation}
\begin{split}
\det{[\lambda^{(1)} + u^{-1}\lambda^{(2)}u]} & = \kappa_1^{(1)}\kappa_2^{(1)} + \kappa_1^{(2)}\kappa_2^{(2)}
+ \sin^2{(\phi)} ( \kappa_1^{(1)}\kappa_1^{(2)} + \kappa_2^{(1)}\kappa_2^{(2)} )\\ 
&\qquad  + \cos^2{(\phi)} ( \kappa_1^{(1)}\kappa_2^{(2)} + \kappa_1^{(2)}\kappa_2^{(1)} )
\end{split}
\end{equation}
and integrating (\ref{glat}) over all relative orientations of the two particles, yields the surface of the Weyl tube:
\begin{equation}\label{mink1}
\int_{\Lambda_{ij}} g_{ij}^\H\, d\gamma_i \approx 8\pi^2 c \int e_{ij}^\H(t) \delta(t) \,dt\, d\sigma_i 
= 8\pi^2 c a_i \delta_{ij}
\end{equation}
and the surface of their Minkowski sum: 
\begin{equation}\label{mink2}
\int_{\Lambda_{ij}} g_{ij}^\H d\gamma_{ij} \approx c \int \det{[\lambda^{(1)} + u^{-1}\lambda^{(2)}u]} d\phi d\sigma_i 
d\sigma_j = 8\pi^2 c (a_i + a_j + \frac{1}{4\pi} \bar\kappa^{(i)}\bar\kappa^{(j)})\;.
\end{equation}

The Minkowski surface is therefore not simply the sum of its individual surfaces but corrected by the product of 
mean curvatures $\bar\kappa^{(i)}$. Its counterpart in the lattice representation are cell segments adjoined at the 
edges of the molecule but not on its surface segments. These cells, however, are ignored in the next-neighbor 
approximation, explaining why the lattice models cannot represent the specific geometry of a particle. Omitting the 
non-additive part and introducing the packing fraction $\eta = \sum_i \rho_i v_i$, determines the remaining three groups 
of variables 
\begin{equation}\label{def-a}
\phi_i = \frac{x_iv_i}{\sum_k x_kv_k}\;,\qquad \theta_i = \frac{x_ia_i}{\sum_k x_ka_k}\;,\qquad z = z_0 \sum_k x_ka_k
\end{equation}
as a function of the universal model parameter $z_0$.

The last, but subtle, step in determining the thermodynamic equilibrium is the minimization of the functional
\begin{equation}\label{var-f}
\delta \Omega^\D = \frac{\delta \Omega^\D}{\delta \rho_k}\Big|_{g_2} \delta \rho_k + \frac{1}{2} \frac{\delta 
\Omega^\D}{\delta 
g_{ij}} \frac{\delta g_{ij}}{\delta \rho_k} \delta \rho_k = 0\;.
\end{equation}
The Euler-Lagrange equation of the first term defines the chemical potential, while the second reproduces the 
constraint (\ref{scf}) as a self-consistent equation. To compare this equation to its analogue in $\Omega^\E$, we apply 
the previous approximations by omitting terms of $g_2$ beyond the linear order $\ln{(g_2)} = -\beta \phi + d_2 + h_2 - 
c_2 \approx -\beta \phi$ and rewrite the correlation function in the basis of the lattice variables (\ref{ident})
\begin{equation}\label{g-lat}
g_{ij}|_{\Lambda_{ij}} = \frac{\theta_{ij}}{\theta_i}\frac{z_i\,z_j}{z} \approx \exp{(-\beta \phi_{ij})}|_{\Lambda_{ij}}
= \exp{(-\beta\epsilon_{ij})}\;.
\end{equation}

The corresponding minimization of $\Omega^\E$ with respect to $\theta_{ij}$ and the constraints (\ref{norm}) yield the 
Euler-Lagrange equation for the lattice model \cite{gug1}
\begin{equation}\label{min-l}
\frac{\delta \Omega^\E}{\delta \theta_{ij}} = 0\;:\qquad 
\frac{\theta_{ij}\theta_{ji}}{\theta_{ii}\theta_{jj}}
= \exp{(-\beta[\,2\epsilon_{ij} -\epsilon_{ii} -\epsilon_{jj}\,])}
\end{equation}
for which we introduce the notations:
\begin{equation}\label{def-t}
\tau_{ij}^2 := \exp{(-\beta[\,2\epsilon_{ij} -\epsilon_{ii} -\epsilon_{jj}\,])}\;,\quad
t_{ij} := \exp{(-\beta[\epsilon_{ij} - \epsilon_{jj}])}\;,\quad \tau_{ij}^2 = t_{ij}\,t_{ji}\;.
\end{equation}
By inserting (\ref{g-lat}) into $g_{ij}g_{ji}/(g_{ii}g_{jj})$, it is easily seen that the minimum of the continuum 
functional and that of its lattice counterpart (\ref{min-l}) agree, therefore proving that the first-shell 
approximation does not violate the thermodynamic consistency of the functional.

In the literature, two different solutions for (\ref{min-l}) can be found. The first one, developed by Larsen and 
Rasmussen (LR) \cite{lars-ras}, uses the symmetry properties (\ref{norm}) to derive the square root
\begin{equation}\label{lr}
\Big(\frac{\theta_{ij}}{\theta_j}\Big)^2 = \frac{\theta_{ii}}{\theta_i}\frac{\theta_{jj}}{\theta_j}\, \tau_{ij}^2\;, 
\quad b_i^2 := \frac{\theta_{ii}}{\theta_i} \quad \Rightarrow \quad \frac{1}{b_j} = \sum_i \tau_{ij}\, \theta_i\, b_i\;,
\end{equation}
which can be numerically solved for $b_i$ and back-inserted to obtain $\theta_{ij}$. The alternative approach goes back 
to Wilson (W) \cite{wilson} and uses the ad hoc separation
\begin{equation}\label{w}
\frac{\theta_{ij}}{\theta_{jj}} = \frac{\theta_i}{\theta_j}\, t_{ij}\;,\quad
\frac{\theta_{ji}}{\theta_{ii}} = \frac{\theta_j}{\theta_i}\, t_{ji}\qquad \Rightarrow \qquad 
\theta_{ij} = \frac{\theta_i\, t_{ij}}{\sum_k \theta_k\, t_{kj}}\;,\quad
\theta_{ji} = \frac{\theta_j\, t_{ji}}{\sum_k \theta_k\, t_{ki}}
\end{equation}
to obtain two independent solutions for (\ref{min-l}) in terms of $t_{ij}$. This approach, however, is inconsistent, as 
can be seen from the missing permutation invariance of $t_{ij}$ in its indices and by inserting (\ref{g-lat}) into 
$g_{ij}/g_{jj}$:
\begin{equation}
\frac{\theta_{ij}}{\theta_{jj}} = \frac{\theta_i}{\theta_j}\,\frac{z_j}{z_i}\,t_{ij} = \frac{\phi_i}{\phi_j}\,t_{ij}\;.
\end{equation}
The Wilson ansatz is therefore only a formal solution, depending either on the volume or the surface fraction and at 
most applicable for molecules of similar spherical size $z_i \approx z_j$.

Inserting (\ref{lr}), (\ref{w}) into the functional (\ref{uni}) and taking account of the two independent solutions of 
the Wilson model, yields the minimum of the excess free-energy with respect to $\theta_{ij}$
\begin{align}
\beta F_\text{LR}^\E/N & = \sum_i x_i\ln{(\phi_i)} + \frac{z}{2}\sum_i\theta_i \ln{\Big(\frac{\theta_i}{\phi_i}\Big)}
+ \frac{1}{2} zN \sum_i \theta_i \ln{\Big[\frac{\theta_{ii}}{\theta_i} \Big]}\;,\label{lrf}\\
\beta F_\text{\,W}^\E/N & = \sum_i x_i\ln{(\phi_i)} + \frac{z}{2}\sum_i\theta_i 
\ln{\Big(\frac{\theta_i}{\phi_i}\Big)} - zN \sum_i \theta_i \ln{\Big[\sum_j \theta_j\, t_{ji} \Big]}\;,\label{uniquac}
\end{align}
where the second result corresponds to the UNIQUAC model introduced by Prausnitz, Abrams, and Maurer 
\cite{praus1,praus2}. 

The liquid-liquid equilibrium at a given reference point of density and pressure is now determined by the excess 
free-energy function and the parameters $v_i$, $a_i$, $\tau_{ij}$ and $t_{ij}$ respectively. Their values can be 
adjusted to experimental data if a sufficiently large set is known. This is especially convenient for the analytical 
solution of the Wilson ansatz (\ref{w}), which partly explains the popularity of the UNIQUAC model. If, however, the 
data set is too small, one has to resort to further models to specify the geometry and intermolecular potentials. One 
such approach is the group-contribution approximation, which uses the observation that the chemical and physical 
properties of organic compounds are often dominated by their functional groups. Together with the lattice assumption of 
next-neighbor interactions, the potential $\phi_{ij}$ is replaced by a superposition of interactions 
$\phi_{\alpha\beta}$ of its $\alpha, \beta = 1, \ldots, N_G$ functional groups
\begin{equation}\label{group-pot}
\phi_{ij} = \sum_{\alpha\beta}\, n_i^\alpha \, n_j^\beta \, \phi_{\alpha\beta}\;,
\end{equation}
related to an analogous transformation of the pair-correlation functionals
\begin{equation}\label{group}
\frac{\delta\Omega}{\delta \phi_{ij}} = \sum_{\alpha\beta} \frac{\delta\Omega}{\delta \phi_{\alpha\beta}} n_\alpha^i 
n_\beta^j\;:\quad
\rho_{ij} = \sum_{\alpha\beta} \rho_i \rho_j n_\alpha^i n_\beta^j g_{\alpha\beta}\;.
\end{equation}

The functional groups are the lattice equivalent of the site-site interactions used for molecular fluids 
\cite{mcdonald}. But in combination with the next-neighbor approach, they decouple and formally replace the molecules 
as individual particles in the potential part of the free energy. Writing its contribution in group indices, the 
transformation leaves the particle density $\rho_i$ and the product of canonically conjugate variables $g_{ij}\phi_{ij} 
= g_{\alpha\beta} \phi_{\alpha\beta}$ invariant. Only the integral measure $d\gamma_{ij} = n_i^\alpha n_j^\beta 
d\gamma_{\alpha\beta}$ is changed by the transition $d\sigma_i =|d\sigma_i/d q_\alpha| d q_\alpha$ from surface elements 
to surface groups or charges $q_\alpha$. The transformation of the potential energy thus remains formally invariant
\begin{equation}
\sum_{ij} \int \rho_i\rho_j g_{ij} \phi_{ij} d\gamma_{ij} = \sum_{ij}\sum_{\alpha\beta} \int \rho_i \rho_j 
g_{\alpha\beta} \phi_{\alpha\beta} n_i^\alpha n_j^\beta d\gamma_{\alpha\beta} 
= \sum_{\alpha\beta} \int \rho_\alpha \rho_\beta g_{\alpha\beta} \phi_{\alpha\beta} d\gamma_{\alpha\beta}\;,
\end{equation}
if the particle density is redefined as the density of group elements
\begin{equation}
\rho_\alpha = \sum_i n_i^\alpha \rho_i\;.
\end{equation}

Because the partial integration (\ref{part-trafo}) commutes with the coordinate change (\ref{group-pot}), the same 
transformation applies to the complete functional. The excess free energy (\ref{uni}) and the models (\ref{lrf}), 
(\ref{uniquac}) therefore remain formally invariant. Using the substitution $x_i = n_i^\alpha x_\alpha$ for the molar 
fractions, yields the lattice variables of the group-contribution models
\begin{equation}
\theta_\alpha = \frac{\sum_i x_i n_i^\alpha a_\alpha}{\sum_k x_k a_k}\;, \quad \theta_{\alpha\beta}\;,\quad 
\tau_{\alpha\beta}\;,\quad t_{\alpha\beta}\;,
\end{equation}
for the group surface $a_\alpha = n_\alpha^i a_i$ and the group volume $v_\alpha = n_\alpha^i v_i$.

Writing the UNIQUAC equation in the basis of group contributions reproduces the UNIFAC model \cite{praus3}. Its
extended class of parametrized functional groups improves the accuracy of the UNIQUAC model and allows to interpolate 
between molecules of similar chemical structure. But its dependence on the Wilson ansatz, the low spacial resolution 
of the interaction potential, and the heuristic notion of functional groups limits its value as a guideline for further 
improvements. 

An approach that avoids these complications is the COMOS-RS model \cite{klamt-1,klamt-0}. Instead of the functional 
groups it uses partial charges $q_{i_\alpha}$ localized at the segments $a_{i_\alpha}$ of the discretized surface of 
the molecule. Their values are derived by a quantum mechanical COSMO calculation, approximating the dielectric 
background of the liquid by the boundary condition of a conducting surface, which can be solved by inserting mirror 
charges $-q_{i_\alpha}$ \cite{cosmo1}. Removing the boundary condition, these charges generate an electrical field 
$\vec E$ outside the particle, pointing into the normal direction $\hat n_{i_\alpha}$ of the segments $a_{i_\alpha}$. 
This has to be taken into account, when the interaction energy between two neighboring molecules is determined. Using 
the Maxwell tensor $\sigma_{ab} = 1/(4\pi) (E_aE_b -E^2\delta_{ab}/2)$, the energy of the electric field between the 
surface charges $q_{i_\alpha}$, $q_{j_\beta}$, separated by a distance $t_{i_\alpha j_\beta}$, is approximately
\begin{equation}\label{s-int}
\phi_{i_\alpha j_\beta } = \kappa\; q_{i_\alpha} q_{j_\beta} \frac{\hat n_{i_\alpha} \hat 
n_{j_\beta}}{t_{i_\alpha j_\beta}}\;.
\end{equation}
Inserting this result into (\ref{def-t}), yields the interaction matrix for (\ref{lr})
\begin{equation}\label{ctau}
\tau_{i_\alpha j_\beta} = \exp{[-\frac{\beta}{2}\frac{\kappa}{t_{i_\alpha j_\beta}} 
(q_{i_\alpha}\hat n_{i_\alpha} - q_{j_\beta} \hat n_{j_\beta})^2]}\;.
\end{equation}

Solving the self-consistent equation is still a time-consuming task even for small molecules. Given a mixture of 
particles with $S_i$ surface segments, the rank of the matrix is $S_1 + \ldots + S_M$, which for binary mixtures is 
of order $\sim 10^3-10^4$. To shorten the calculation time, the COSMO-RS model introduces group variables to coarse 
grain the number of charges $q_\alpha = n_\alpha^i q_{i_\alpha}$ and segments $a_\alpha = n_\alpha^i a_{i_\alpha}$, 
simplifying the self-consistent equation \cite{klamt-0}
\begin{equation}\label{space}
\frac{1}{b_\beta} = \sum_\alpha \tau_{\alpha\beta} \theta_\alpha b_\alpha\;,\quad
\tau_{\alpha\beta} = \exp{[-\frac{\beta}{2}\kappa (q_\alpha + q_\beta)^2]}\;,
\end{equation}
for molecules of antiparallel surface segments $\hat n_{i_\alpha}= -\hat n_{j_\beta}$ and separated by an average 
distance $t_{i_\alpha j_\beta}= t_0$, whose value has been absorbed in the overall constant $\kappa$. Together with the 
reference geometry of the unit sphere, this corresponds to the interaction model shown in Fig.~\ref{fig:segs}a). 

Apart from the electrostatic interaction, the COSMO-RS model also includes dispersion effects and hydrogen bonding, but 
fails for the Coulomb interaction. This is to be expected, as the next-neighbor ansatz requires the correlation length 
to be of the order of the first particle shell, whereas the correlation length of strong electrolytes is significantly 
larger. As a result, the ring integrals (\ref{ring}) can no longer be ignored. A first approximation is therefore to 
couple the Debye-H{\"u}ckel (\ref{dh}) or Debye-H{\"u}ckel-Pitzer term to the grand potential \cite{cosmo-sac-e-1, 
cosmo-rs-e-1} and to derive the new closure condition from the Euler-Lagrange equation. For the example of an 
electrolyte with point charges $\pm q$, this yields an implicit equation in $g_2$
\begin{equation}
\ln{(g_2)} = -\beta \phi - F^{-1}\Big( \frac{\rho \hat h_2^2}{1+\rho \hat h_2} \Big)\;.
\end{equation}
Its algebraic solution is no longer possible. But the pair correlation is still dominated by the potential $\phi$ at 
particle contact $r = t_0$ and only modified by the Debye-H{\"u}ckel term at distances $r_0 = 2\pi/k_D$. If $r_0 \approx 
t_0$, the screening effect is small and can therefore be ignored. Whereas the detailed geometry of strong electrolytes 
$r_0 \gg t_0$ is less important for larger distances, and the pair-correlation function $h_2(r_0)$ can be approximated 
by (\ref{debye}). The inhomogeneous charge distribution has then the effect of a charged dielectric background that
contributes a compensating potential to $\phi$ without violating the additive structure required for the 
self-consistent equation of the lattice model (\ref{lr}).

Despite this generalization, lattice models remain limited by the fixed reference value of density and pressure and the 
neglect of the molecular geometry. Any improvement requires the construction of a density functional. A natural link 
between both descriptions is the COSMO model. The cavity and partial charges provide the necessary information to 
define the hard-particle geometry and soft interaction for the approximate pair-correlation functional 
$g_{ij}(\sigma_i, \sigma_j, t_{ij}, \phi)$, as shown in Fig.~\ref{fig:segs}b). It is a function of the surfaces 
$\sigma_i$, $\sigma_j$, separated by the distance $t_{ij}$, and rotated by the axial angle $\phi$. It also introduces 
correlations between spatially separated surface segments, which is unavoidable to describe elongated or concave  
molecules and to approach problems from biology and chemistry.
\section{Discussion and Conclusion}\label{sec:conclusion}
The last two sections have shown that $\Omega^\D$ is a promising starting point for the construction of molecular 
density functionals. It has a simple first perturbation order, a local consistency test for the pair-correlation 
functional, and it reproduces the excess free energy of the lattice models. By combining the various approximation 
methods, we have now a better understanding for the continuum functional and how to combine the quantum mechanical data 
of a COSMO calculation with the simple interaction model of classical mechanics.

On the other hand, the derivation also gives new insight into the structure of the lattice models. We have shown that 
the self-consistent identity follows from to the Euler-Lagrange equation of the pair-correlation functional. This adds 
a third route for its derivation to the previous approaches developed by Klamt and Sandler. And it also shows that its 
solution by the Larsen-Rasmussen ansatz corresponds to locating the single minimum of the grand potential, while 
Wilson's algebraic approach only yields an approximate result.

Apart from the three investigates models, there exists several more modifications and realizations that have not been 
discussed here. Most often, they differ in their approach to solve the self-consistent equation or to couple further 
interactions, with the Coulomb potential as the most relevant example. Including this long-range interaction, adds the 
Debye-H{\"u}ckel-Pitzer term to the grand potential, associated with a corresponding change of the Euler-Lagrange 
equation. A consistent implementation of the Coulomb interaction therefore has to modify the energy functional as well 
as the self-consistent equation of the lattice.

The main advantage of the lattice models is their computational efficiency. Solving the self-consistent equation of the 
COSMO-RS model only takes seconds. Whereas the calculation time in the basis of surface segments is of the order of 
hours. And the minimization of the density functional will take even longer. It is therefore necessary to develope 
further approximations for the hard-particle correlations and the integration of the grand potential. As for the 
COSMO-RS model, it is possible to assume a fixed average particle distance for the molecular functional and to reduce 
the integration over the Euclidean volume to a sum over all segment pairings and rotations. 

The possibility to derive the thermodynamic equilibrium from a surface integral illustrates the potential advantage of 
the DFT ansatz compared to molecular dynamic and Monte Carlo simulations. These two methods apply to the complete 
phase diagram, allow for flexible atomic bonds, and the implementation of boundary conditions. But it is quite 
difficult to reduce this freedom, when one is only interested in a small interval of the phase diagram or in specific 
aspects of the intermolecular properties. Well known examples are the solubility of proteins, the contact probability 
of enzymes, the miscibility of racemic mixtures, or the selection of an optimal chiral selector in liquid 
chromatography. These are only some examples where the usage of an optimized density functional might prove favorable 
compared to the explicit ensemble averaging of the free energy.
\section*{Acknowledgements}
The author wishes to thank Andr\'{e} Bardow and Kai Leonhard for their support of this work and helpful comments on the 
manuscript.
\appendix
\section{Appendix}\label{sec:app}
The r-particle correlation functionals are always accompanied by an integration over the kinematic measure 
$d\gamma_{i_1\ldots i_r}$ of translations and rotations. For the most common case of $r=2$, we will now derive an 
explicit realization using methods from integral geometry \cite{guggenheimer, weyl-tube, hsiung, minkowski}.

Let $\Sigma_k$ denote a $n-1$ dimensional, smooth, boundary free, convex, Riemannian manifold, imbedded into $D_k: 
\Sigma_k \hookrightarrow \mathbb{R}^n$. Each point $p_k\in \Sigma_k$ is then related to an orthonormal, positively 
oriented coordinate frame $(\hat e_1^{(k)}, \ldots, \hat e_n^{(k)})$ with the outward pointing normal vector $\hat 
e_n^{(k)}$, differential basis $\theta_i$, and connection forms $\omega_{ij}$
\begin{equation}\label{riem}
dp = \theta_i \hat e_i   \;,\quad d\hat e_i = \omega_{ij} \hat e_j\;, \quad d\hat e_n = \omega_{n\alpha}\hat e_\alpha = 
h_{\alpha\beta} \theta_\beta \hat e_\alpha\;,\quad \lambda_{\alpha\beta}:= h_{\alpha\beta}^{-1}\;.
\end{equation}

Each point of the smooth surface is uniquely related to a tangential plane up to an axial rotation around $\hat 
e_n^{(k)}$. From this follows that the tangential planes of two convex surfaces $\Sigma_1$, $\Sigma_2$, touching in a 
common point $p_1 = p_2$, also agree up to an axial rotation and an inversion of their normal vectors $\hat e_n^{(1)} = 
- \hat e_n^{(2)}$. This property remains unchanged even when the particles are shifted apart in the normal direction. 
Surface points of closest distance $t\in\mathbb{R}^+$ and their frames are therefore related by
\begin{equation}\label{coord}
p_2 = p_1 + t \hat e_n^{(1)}\;,\quad \hat e_n^{(1)} = -\hat e_n^{(2)}\;,\quad \hat e_\alpha^{(1)} = u_{\alpha\beta} 
\hat e_\beta^{(2)}\quad \text{for}\quad u_{\alpha\beta}\in\text{SO}(n-1)
\end{equation}
using the index conventions $i,j = 1,\ldots, n$ and $\alpha, \beta = 1,\ldots, n-1$. 

Having defined the relative coordinate frames for the two particles, we can now write the integral
\begin{equation}
\int g_{12}(\vec r_1, \vec r_2)\, d\gamma_1 d\gamma_2 = \int g_{12}(\vec r_1 -\vec r_2)\,d\widetilde\gamma_{12} d\gamma 
= V\text{vol}(\text{SO}(n)) \int g_{12}^\H(\vec r_1 -\vec r_2) \,d\widetilde\gamma_{12}
\end{equation}
in a comoving $d\widetilde\gamma_{12}$ and a reference system $d\gamma$. The integral of the latter can be carried out, 
contributing the volume $V = \text{vol}(\mathbb{R}^n)$ and the volume of the group $\text{SO}(n)$. The analogous 
transformation in the representation of base forms corresponds to the shift
\begin{equation}\label{kin}
d\gamma_1 d\gamma_2 =  \bigwedge_{k=1,2} \bigwedge_i \theta_i^{(k)} \bigwedge_{i<j} \omega_{ij}^{(k)}
= \bigwedge_i(\theta_i^{(1)} - \theta_i^{(2)}) \bigwedge_{i<j} \omega_{ij}^{(1)} \bigwedge_i \theta_i^{(2)} 
\bigwedge_{i<j} \omega_{ij}^{(2)} = d\widetilde\gamma_{12} d\gamma
\end{equation}
as can be seen by expanding the skew-symmetric product and setting $d\gamma = d\gamma_2$ for the reference system.

The trivial contribution $d\gamma$ will be ignored in the following, leaving us with the transformation of $d\widetilde 
\gamma_{12}$. To simplify the calculation, observe that the translation of $\Sigma_2$ can also be written as $p_2 + 
t_2\hat e_2^{(2)} = p_1 + t_1\hat e_n^{(1)}$ for any $t_1,t_2 \in \mathbb{R}^+$ and $t_1+t_2 = t$. This allows to first
determine Weyl's half-tube surface $\Sigma(t): p(t) = p + t\hat e_n$ for $\Sigma_1$, $\Sigma_2$ separately and then to 
derive their Minkowski sum $\Sigma_1(t_1)\oplus \Sigma_2(t_2)$ at $p_1(t_1) = p_2(t_2)$ \cite{weyl-tube, minkowski}. 

In the first step, we determine the differential forms $\theta_i(t), \omega_{ij}(t)$ of the half-tube $\Sigma(t)$ by 
differentiating each point $p(t) = p + t\hat e_n$
\begin{equation}
dp(t) = dp + \hat e_n dt + t\omega_{n\alpha}\wedge \hat e_\alpha
\end{equation}
and separating their components into the directions of $\hat e_n$ and $\hat e_\alpha$
\begin{equation}\label{theta} 
\theta_n(t) = \theta_n + dt\;,\quad \theta_\alpha(t) = \theta_\alpha + t\omega_{n\alpha}\wedge \theta_\alpha
= (\delta_{\alpha\beta} + t h_{\alpha\beta})\theta_\beta\;.
\end{equation}
The new basis also determines the connection forms, as the orthonormal vectors $\hat e_\alpha$ and their differentials 
are invariant under translations
\begin{equation}
\omega_{n\alpha}(t) = \omega_{n\alpha}\quad \Rightarrow \quad h_{\alpha\beta}\theta_\beta(t) = 
h_{\alpha\beta}\theta_\beta\;.
\end{equation}
Inserting (\ref{theta}), finally yields the curvature matrix and its inverse for the half-tube $\Sigma(t)$
\begin{equation}\label{wein}
h_{\alpha\beta}(t) = h_{\alpha\gamma}\,(\delta_{\gamma\beta} + t h_{\gamma\beta})^{-1}\;,\quad 
\lambda_{\alpha\beta}(t) = t\,\delta_{\alpha\beta} + \lambda_{\alpha\beta}
\end{equation}

The second step requires to determine the differential volume element for the domain $\Sigma_1(t_1)\oplus 
\Sigma_2(t_2)$, covered by $\Sigma_1(t_1)$ while circling $\Sigma_2(t_2)$. To write $d\widetilde\gamma_{12}$ in a 
common coordinate frame we use the transformation (\ref{coord}), which relates the connection forms 
$\omega_{n\alpha}^{(1)} = -u_{\alpha\beta} \omega_{n\beta}^{(2)}$ of the two particles at their intersection point 
$p_1(t_1) = p_2(t_2)$ and also defines the transformation of their basis forms. Using $\Sigma_2(t_2)$ as reference 
system, the forms of $\Sigma_1(t_1)$ are rotated into the new coordinate frame
\begin{equation}
\theta_\alpha^{(2)} = \lambda_{\alpha\beta}^{(2)} \omega_{n\beta}^{(2)}\;,\quad 
\theta_\alpha^{(1)} =  u_{\alpha\beta}\theta_\beta^{'(1)} = - u_{\alpha\beta}\lambda_{\beta\gamma}^{(1)} 
u_{\gamma\mu}\omega_{n\mu}^{(2)}\;.
\end{equation}

Inserting this result into (\ref{kin}), together with the transformation of the normal component $\theta_n^{(1)} - 
\theta_n^{(2)} = -dt$ and the Jacobi determinant $J=-1$, yields the reduced kinematic measure
\begin{equation}\label{g-12}
\begin{split}
d\widetilde\gamma_{12} & = \bigwedge_{i}(\theta_i^{(1)} - \theta_i^{(2)}) \bigwedge_{i<j}\omega_{ij}^{(1)}\\
& = \bigwedge_\alpha ( u\lambda^{(1)}(t_1)u^{-1} +  \lambda^{(2)}(t_2))_{\alpha\beta}\, \omega_{n\beta}^{(2)} \wedge 
dt\bigwedge_\alpha \omega_{n\alpha}^{(1)}\bigwedge_{\alpha <\beta} \omega_{\alpha\beta}^{(1)} \\
& = \det{(\lambda^{(1)} + t\delta + u^{-1}\lambda^{(2)}u)}\, \kappa_G^{(1)}\kappa_G^{(2)} \, d\sigma_1 \wedge 
d\sigma_2\wedge d\text{SO}(n-1)\wedge dt\;,
\end{split}
\end{equation}
where we introduced the unit matrix $\delta$, the Gaussian curvature $\wedge_\alpha \omega_{n\alpha} = \kappa_G\, 
d\sigma$, the differential surface element $d\sigma$, used the orthonormal property $\det{(u)} = 1$ and (\ref{wein}) to 
write the final result in a symmetric form.
\bibliographystyle{tMPH}
\bibliography{molecular_dft}

\newcommand{\noopsort}[1]{} \newcommand{\printfirst}[2]{#1}
  \newcommand{\singleletter}[1]{#1} \newcommand{\switchargs}[2]{#2#1}
\begin{thebibliography}{41}
\providecommand{\url}[1]{\texttt{#1}}
\providecommand{\urlprefix}{URL }
\markboth{Taylor \& Francis and I.T. Consultant}{Molecular Physics}

\bibitem{mcdonald}
I.R. McDonald and J.P. Hansen, \emph{Theory of Simple Liquids}   (Academic
  Press, Burlington, 2013).

\bibitem{evans-dft}
R. Evans,  Adv. Phys.  \textbf{28}, 143 (1979).

\bibitem{korden-2}
S. Korden,  Phys. Rev. E  \textbf{85} (4), 041150 (2012).

\bibitem{korden-4}
S. Korden, Density Functional Theory for Hard Particles in N Dimensions,
  arXiv:1403.2054, accepted by Comm. Math. Phys. (unpublished) 2015.

\bibitem{korden-6}
S. Korden, Distribution Functionals for Hard Particles in N Dimensions,
  arXiv:1502.04393 (unpublished) 2015.

\bibitem{klamt-0}
A. Klamt, \emph{COSMO-RS: From Quantum Chemistry to Fluid Phase Thermodynamics
  and Drug Design}   (Elsevier Science, Amsterdam, 2005).

\bibitem{praus4}
J. Prausnitz, R. Lichtenthaler and E.G. de~Azevedo, \emph{Molecular
  Thermodynamics of Fluid-Phase Equilibria}   (Prentice-Hall, Englewood Cliff,
  NJ, 1999).

\bibitem{prigog1}
I. Prigogine, \emph{The Molecular Theory of Solutions}   (North-Holland Pub.
  Co., Amsterdam, 1957).

\bibitem{gub-rev}
K.E. Gubbins,  Molecular Physics  \textbf{111}, 3666 (2013).

\bibitem{hill}
T.L. Hill, \emph{Statistical Mechanics: Principles and Selected Applications}
  (Dover Publications, New York, 1956).

\bibitem{baxter2}
J. Baxter, \emph{Lattice Theories of the Liquid State}   (Pergamon Press,
  Oxford, 1963).

\bibitem{flory1}
P. Flory,  J. Chem. Phys.  \textbf{9}, 660 (1941).

\bibitem{flory2}
P. Flory,  J. Chem. Phys.  \textbf{10}, 51 (1942).

\bibitem{huggins}
M. Huggins,  J. Chem. Phys.  \textbf{9}, 440 (1941).

\bibitem{gug1}
E. Guggenheim, \emph{Mixtures}   (Clarendon Press, Oxford, 1952).

\bibitem{praus1}
G. Maurer and J. Prausnitz,  Fluid Phase Equilibria  \textbf{2}, 91 (1978).

\bibitem{praus2}
D.S. Abrams and J.M. Prausnitz,  AIChE Journal  \textbf{21}, 116 (1975).

\bibitem{praus3}
A. Fredenslund, R.L. Jones and J.M. Prausnitz,  AIChE Journal  \textbf{21},
  1086 (1975).

\bibitem{klamt-1}
A. Klamt,  J. Phys. Chem.  \textbf{99}, 2224 (1995).

\bibitem{klamt-2}
A. Klamt, F. Eckert and W. Arlt,  Annu. Rev. Chem. Biomol. Eng.  \textbf{1},
  101 (2010).

\bibitem{sac_1}
S.T. Lin and S.I. Sandler,  Ind. Eng. Chem. Res.  \textbf{41}, 899 (2002).

\bibitem{sac_2}
C.M. Hsieh, S.I. Sandler and S.T. Lin,  Fluid Phase Equilibria  \textbf{297},
  90 (2010).

\bibitem{morita3}
T. Morita and K. Hiroike,  Prog. Theor. Phys.  \textbf{25}, 537 (1961).

\bibitem{baxter3}
R. Baxter, in \emph{Physical Chemistry, An Advanced Treatise}, edited by D.
  Henderson  (Academic Press, New York, 1971), Vol. VIII A, pp. 267--334.

\bibitem{morita1}
T. Morita and K. Hiroike,  Prog. Theor. Phys.  \textbf{23}, 1003 (1960).

\bibitem{morita2}
T. Morita and K. Hiroike,  Prog. Theor. Phys.  \textbf{24}, 317 (1960).

\bibitem{morita4}
T. Morita and K. Hiroike,  Prog. Theor. Phys.  \textbf{25}, 537 (1961).

\bibitem{wertheim-py1}
M.S. Wertheim,  Phys. Rev. Lett.  \textbf{10}, 321 (1963).

\bibitem{wertheim-py2}
M.S. Wertheim,  J. Math. Phys.  \textbf{5}, 643 (1964).

\bibitem{thiele}
E. Thiele,  J. Chem. Phys.  \textbf{39}, 474 (1963).

\bibitem{baxter}
R. Baxter,  Aust. J. Phys.  \textbf{21}, 563 (1968).

\bibitem{pitzer}
K.S. Pitzer,  Acc. Chem. Res.  \textbf{10}, 371 (1977).

\bibitem{lars-ras}
B. Larsen and P. Rasmussen,  Fluid Phase Equilibria  \textbf{28}, 1 (1986).

\bibitem{wilson}
G. Wilson,  J. Am. Chem. Soc.  \textbf{86}, 127 (1964).

\bibitem{cosmo1}
A. Klamt and G. Sch{\"u}{\"u}rmann,  J. Chem. Soc., Perkin Trans.  \textbf{2},
  799 (1993).

\bibitem{cosmo-sac-e-1}
S. Wang, Y. Song and C.C. Chen,  Ind. Eng. Chem. Res.  \textbf{50}, 176 (2011).

\bibitem{cosmo-rs-e-1}
T. Ingram, T. Gerlach, T. Mehling and I. Smirnova,  Fluid Phase Equilibria
  \textbf{314}, 29 (2012).

\bibitem{weyl-tube}
H. Weyl,  Am. J. Math.  \textbf{61}, 461 (1939).

\bibitem{guggenheimer}
H.W. Guggenheimer, \emph{Differential Geometry}   (Dover Publications, New
  York, 1963).

\bibitem{hsiung}
C.C. Hsiung, \emph{A First Course in Differential Geometry}   (Int. Press
  Boston Inc., USA, 2013).

\bibitem{minkowski}
H. Minkowski,  Math. Ann.  \textbf{57}, 447 (1903).

\end{thebibliography}
\end{document}